\documentclass[twocolumn,amsmath,aps]{revtex4}
\usepackage{graphicx}

\newcommand{\nl}{\nonumber \\}

\newcommand{\be}{\begin{equation}}
\newcommand{\ee}{\end{equation}}
\newcommand{\bea}{\begin{eqnarray}}
\newcommand{\eea}{\end{eqnarray}}

\newcommand{\Eq}[1]{Eq.\,(\ref{#1})}
\newcommand{\Eqs}[1]{Eqs.\,(\ref{#1})}

\newcommand{\la}{\langle}
\newcommand{\ra}{\rangle}
\newcommand{\dg}{\dagger}

\newcommand{\ti}{\tilde}
\newcommand{\mb}{\mbox}

\begin{document}
\draft

\title{Quantum measurement of an electron in disordered potential:
       delocalization versus measurement voltages }
\author{Xue-Ning Hu and Xin-Qi Li }
\address{State Key Laboratory for Superlattices and Microstructures,
         Institute of Semiconductors,
         Chinese Academy of Sciences, P.O.~Box 912, Beijing 100083, China }

\date{\today}

\begin{abstract}
Quantum point contact (QPC), one of the typical mesoscopic
transport devices, has been suggested to be an efficient detector
for quantum measurement. In the context of two-state charge qubit,
our previous studies showed that the QPC's measurement back-action
cannot be described by the conventional Lindblad quantum master
equation.
In this work, we study the measurement problem of a multi-state
system, say, an electron in disordered potential, subject to the
quantum measurement of the mesoscopic detector QPC. The effect of
measurement back-action and the detector's readout current are
analyzed, where particular attention is focused on some new
features and the underlying physics associated with the
measurement-induced delocalization versus the measurement
voltages.\\
\\
\\
\\
PACS numbers: 03.67.Lx, 73.23.-b, 85.35.Be
\end{abstract}
\maketitle

{\it Introduction.}--- Measurement of an individual quantum state
in solid-state system has attracted wide spread attention in
recent years, largely due to the promising field of solid-state
quantum computation. A possible solid-state implementation of such
measurement is to measure a charge qubit by a mesoscopic quantum
point contact (QPC) \cite{Gur97,Win97}.
Very recently, an elegant experiment was performed by employing a
QPC to measure the quantum dot occupation by an extra electron,
which is further associated with a single electron spin state
\cite{Elz04}. This experiment clearly demonstrated the extremely
high sensitivity of the QPC detector, implying its possible wide
application in the future. It is therefore of importance to
develop reliable theoretical description for this important quantum
measurement device.

Theoretically, this measurement problem was first studied by
Gurvitz \cite{Gur97}, followed then by many other groups
\cite{Win97}. Here we mention three typical approaches employed in
literatures: (i) the so-called Bloch equation approach developed
in Ref.\ \onlinecite{Gur97} and a number of other papers by
Gurvitz {\it et al}; (ii) the quantum trajectory technique from
quantum optics by Goan {\it et al} \cite{Goa01}; and (iii) the
Bayesian approach by Korotkov {\it et al} \cite{Kor99}.
In spite of their different forms in appearance, these three
approaches are equivalent in essence.
In particular, all of them are based on
the same (unconditional) Lindblad master equation.
However, as clearly manifested in Ref.\ \onlinecite{Gur97},
the associated Lindblad master equation would result
in the {\it universal} equal occupation probability
on the qubit states in stationary state.
Obviously, under finite voltages this result breaks down the detailed
balance condition, which is thus valid only at high voltage limit.

Very recently, in the context of two-state quantum measurement, we
showed that in general (i.e. under arbitrary measurement
voltages) the QPC measurement setup cannot be described by the
Lindblad master equation \cite{Li04,Li05}.
In this work,
based on our new treatment we study the quantum measurement of a
{\it multi-state} system, say, an electron in disordered potential,
measured by the mesoscopic detector QPC.
It is well known that an electron in disordered potential will be
subject to the Anderson localization, and {\it delocalization}
will take place under quantum measurement
\cite{Dit90,Fac99,Flo99}.
In particular, without using any strong projection postulate,
Gurvitz showed that {\it complete delocalization}, i.e., absolute
equal occupation probabilities on each local state, will be
inevitably approached, under the influence of local measurement by
a QPC \cite{Gur00}. There, the delocalization was also interpreted
in terms of the measurement-induced loss of quantum coherence.
In this work, we show that the delocalization behavior
predicted in Ref.\ \onlinecite{Gur00} is valid only
at high voltage limit, and extend the study to arbitrarily
finite voltages.
Also, we conclude that the delocalization is largely a result of energy
exchange of the measured electron with the measurement device QPC,
thus depends on the measurement voltages.


{\it Model Description.}--- As schematically shown in Fig.\ 1,
consider an electron in a one-dimensional array of coupled quantum
wells, which is measured by a mesoscopic QPC.
The entire system Hamiltonian reads
\begin{subequations} \label{H1}
\begin{eqnarray}
H   &=&  H_0+H'  ,
\\
H_0 &=&  H_{S}   +  \sum_k (\epsilon^L_k c^{\dg}_kc_k+\epsilon^R_k
d^{\dg}_kd_k) ,
\\
H_{S} &=& \sum_{j=1}^N \epsilon_jc^{\dg}_jc_j+\sum^{N-1}_{j=1}
(\Omega_jc^{\dg}_{j+1}c_j+\mb{H.c.}) ,
\\
H'  &=& \sum_{k,q} (\Omega_{qk}+\sum_j \chi_{qk}^jc^{\dg}_jc_j)
         c^{\dg}_k d_q  + \mb{H.c.} .
\end{eqnarray}
\end{subequations}
In this decomposition, the free part of the total Hamiltonian,
$H_0$, contains Hamiltonians  of the measured system ($H_S$) and
the QPC reservoirs (the last two terms). The operator
$c_j^{\dagger}(c_j)$ corresponds to the creation (annihilation) of
an electron in the $j_{\rm th}$ well. For simplicity we assume
that each well contains a single bound state $\epsilon_j $ and is
coupled only to its nearest neighbors with couplings $\Omega_j$
and $\Omega_{j-1}$.
$c_k^{\dag}(c_k)$ and $d_k^{\dag}(d_k)$ are, respectively,
the electron creation (annihilation) operators of the left
and right reservoirs of the QPC.
The Hamiltonian $H'$ describes electron tunneling
between the two reservoirs of the QPC detector
with, for instance, tunneling amplitude $\Omega_{qk}+\sum_j
\chi_{qk}^jc^{\dg}_jc_j$, which generally depends on the measured
electron's position that is characterized by occupation operator
$c^{\dg}_jc_j$. This dependence properly describes the correlation
between the detector and the measured system, which enables to
draw out measurement information from the output current, and
simultaneously propagates back-action of the detector onto the
measured system, causing state dephasing and relaxation.
\begin{figure}\label{fig1}
\begin{center}
\includegraphics*[width=8.5cm,height=9cm,keepaspectratio]{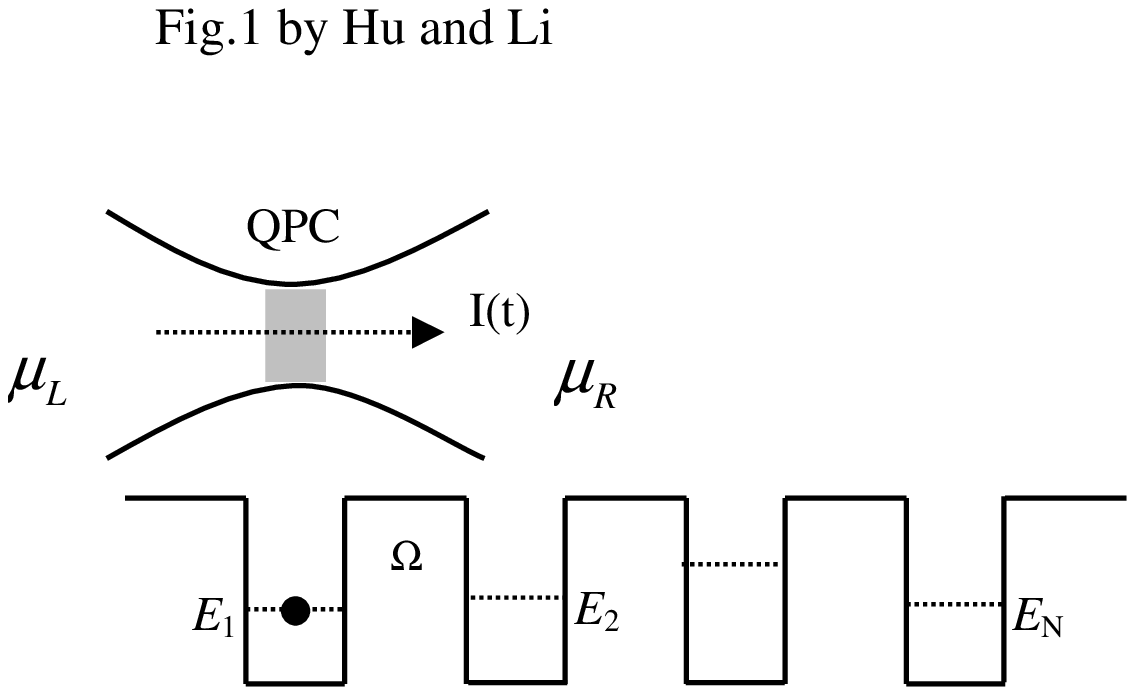}
\caption{ Schematic illustration of using the mesoscopic quantum-point-contact to measure
an electron in multiple coupled quantum wells. }
\end{center}
\end{figure}


{\it Measurement Back-Action}.---
Statistically, the measurement back-action onto the measured
system is described by a quantum master equation (QME) that is
satisfied by the reduced density matrix. Regarding the tunneling
Hamiltonian $H'$ as perturbation, the second-order cumulant
expansion gives rise to a formal equation for the reduced density
matrix \cite{Yan98}
\begin{equation}\label{ME-1}
\dot{\rho}(t)=-i\mathcal{L}\rho(t)-\int_0^t
d\tau\langle\mathcal{L}'(t)\mathcal{G}(t,\tau)\mathcal{L}'(\tau)
{\mathcal{G}}^\dag(t,\tau)\rangle\rho(t).
\end{equation}
Here the Liouvillian superoperators are defined as
$\mathcal{L}(\cdots)$ $\equiv$ $[H_{S,}(\cdots)]$,
 $\mathcal{L}'(\cdots)$ $\equiv$ $[H',(\cdots)]$, and
$\mathcal{G}(t,\tau)(\cdots)$ $\equiv$
$G(t,\tau)(\cdots)G^\dag(t,\tau)$ with $G(t,\tau)$ the usual
propagator (Green's function) associated with $H_S$.
The reduced density matrix $\rho(t)=\mb{Tr}_D[\rho_T(t)]$,
resulting from tracing out all the detector degrees of freedom
from the entire density matrix.
However, for quantum measurement where the specific readout
information is likely to be recorded, the average should be
performed over the unique class of states of the detector we are
trying to keep track of.

For the measurement setup under study, the relevant quantity of
readout is the transport current $i(t)$ in the detector, or
equivalently, the number of electrons that have tunnelled through
the detector, $n(t)=\int^{t}_{0} dt' i(t')$.
We therefore classify the Hilbert space of the detector as
follows. First, we define the subspace in the absence of electron
tunneling through the detector as ${\cal D}^{(0)}$, which is
spanned by the product of all many-particle states of the two
isolated reservoirs, formally denoted as ${\cal
D}^{(0)}\equiv\mb{span}\{|\Psi_L\ra\otimes |\Psi_R\ra \}$.
Then, we introduce the tunneling operator $f^{\dg}\sim
f^{\dg}_{qk}=d_q^{\dg}c_k$, and denote the Hilbert subspace
corresponding to $n$-electrons tunnelled from the left to the
right reservoirs as ${\cal D}^{(n)}=(f^{\dg})^n {\cal D}^{(0)}$,
where $n=1,2,\cdots$. The entire Hilbert space of the detector is
${\cal D}=\oplus_n{\cal D}^{(n)}$.

With the above classification of the detector states, the average
over states in ${\cal D}$ in \Eq{ME-1} is replaced with states in
the subspace ${\cal D}^{(n)}$. Following Ref.\ \onlinecite{Li05},
a {\it conditional} master equation can be derived as
\bea\label{ME-2} \dot{\rho}^{(n)}
   &=&  -i {\cal L}\rho^{(n)} - \frac{1}{2}
        \left\{  [Q\ti{Q}\rho^{(n)}+\mb{H.c.}] \right. \nl
   & &   - [\ti{Q}^{(-)}\rho^{(n-1)}Q+\mb{H.c.}]   \nl
   & &      \left.  - [\ti{Q}^{(+)}\rho^{(n+1)}Q+\mb{H.c.}] \right\} .
\eea Here $Q\equiv\Omega_0+\sum_j \chi_jc^{\dg}_{j}c_j$. For
simplicity, we have assumed $\Omega_{qk}=\Omega_0$ and
$\chi_{qk}^j=\chi_j$, i.e., the tunneling amplitudes are of
reservoir-state independence.
Other quantities and notations in \Eq{ME-2} are explained as follows.
$\ti{Q}=\ti{Q}^{(+)}+\ti{Q}^{(-)}$, and
$\ti{Q}^{(\pm)}=\ti{C}^{(\pm)}({\cal L})Q$,
with $\ti{C}^{(\pm)}({\cal L})$ the spectral function of the QPC reservoirs.
Under wide-band approximation,
$\ti{C}^{(\pm)}({\cal L})$ can be explicitly
carried out as \cite{Li04}: $\ti{C}^{(\pm)}({\cal L})
  =  \eta \left[x/(1-e^{-x/T}) \right]_{x=-{\cal L}\mp V}$,
where $\eta=2\pi g_Lg_R$, with $g_{L(R)}$ the density of states of
the left (right) reservoir, and $T$ is the temperature. In this
work we use the unit system of $\hbar=e=k_B=1$.
The meaning of the super-operator function $\ti{C}^{(\pm)}({\cal L})$
will become more clear by explicitly carrying out the matrix elements
of $\ti{Q}^{(\pm)}$.
In the eigen-state basis $\{|E_m\ra\}$, we easily obtain
$\ti{Q}^{(\pm)}_{mn}=\ti{C}^{(\pm)}(\pm \omega_{mn})Q_{mn}$,
where $\omega_{mn}=E_m-E_n$ and $Q_{mn}=\la E_m|Q|E_n\ra$.
In this derivation, the simple algebra
$\la E_m|{\cal L}Q|E_n\ra=\la E_m|(H_SQ-QH_S)|E_n\ra
=(E_m-E_n)Q_{mn}$ has been used.
Here we see clearly that the Liouvillian operator
``${\cal L}$" in $\ti{C}^{(\pm)}({\cal L})$
properly involves the energy transfer between the detector and the
measured system into the transition rates,
thus implies a detailed balance condition which determines
the stationary occupation probabilities.

Note that $\rho^{(n)}(t)$ contains rich information about the
measurement. From it, one can obtain the distribution function of
tunnelled electron numbers, the measurement current, and the
output power spectrum, etc.
However, the measurement back-action can be described by a much
simpler equation, i.e., the {\it un-conditional} master equation.
Summing up \Eq{ME-2} over ``$n$", the un-conditional QME satisfied
by the reduced density matrix, $\rho=\sum_n\rho^{(n)}$, is
obtained as
\begin{equation}\label{ME-3}
 \dot{\rho}=-i\mathcal{L}\rho-\frac{1}{2}[Q,\tilde{Q}\rho-\rho\tilde{Q}^\dag] ,
\end{equation}
where the term $[\cdots]$ describes the back-action of the
detector on the measured system.
At high-voltage limit, i.e., the voltage is much larger than the
eigen-energy differences ``$\omega_{mn}$" as mentioned above,
the spectral function $\ti{C}^{(\pm)}({\cal L})\simeq
\ti{C}^{(\pm)}(0)$, and \Eq{ME-3} reduces to a Lindblad-type
master equation \bea\label{LB-ME} \dot{\rho}=-i\mathcal{L}\rho +
\ti{C}(0)\left[ Q\rho Q -\frac{1}{2}(Q^2\rho+\rho Q^2) \right],
\eea where $\ti{C}(0)=\ti{C}^{(+)}(0)+\ti{C}^{(-)}(0)$. It is
straightforward to check that this equation is nothing but the
Lindblad-type master equation used in literature
\cite{Gur97,Goa01,Kor99}.

In the following we first study the measurement induced dephasing
behavior in the {\it low voltage regime}.
Numerically, consider $N=20$ wells with random energy levels ($\epsilon_j$).
The coupling strengths between the nearest-neighbor wells are assumed
identical, $\Omega_j=\Omega$, and the disorder strength $\Delta$
is set to be $\Delta=0.9\Omega$ in the numerical calculation.
For the QPC, we assume $\Omega_{qk}\equiv\Omega_0=\Omega$,
the density of states of the reservoirs $g_L=g_R=0.5/\Omega$,
and the measurement voltage $V=1.5 \Omega$.
From the basics of quantum mechanics, dephasing (or interference
destruction) is resulted from the effort of distinguishing the
superposition components of a quantum state.
Therefore, we adopt two models to reveal the measurement induced
dephasing. One model corresponds to that the experimenter wants to
distinguish the electron's position in each quantum well. This can
be implemented by assuming quantum-well-state-dependent tunneling
coefficients through the QPC in terms of $\Omega+\chi_j$, with
$\chi_j=\Omega/\sqrt{4+(j-1)^{2}}$.
Another is the model studied in Ref.\ \onlinecite{Gur00}, in which
the experimenter only distinguishes the electron in the first
(nearest) well from others, by simply assuming $\chi_j=\Omega/2$
for occupation of the electron in the first well, and $\chi_j=0$
for other occupations.
The resultant faster dephasing behavior from the former model is
plotted in Fig.\ 2(a) and (b), and the slower dephasing from the
latter is shown in Fig.\ 2(c) and (d). As expected, these quite
different dephasing rates are consistent with the general quantum
measurement principle, i.e., stronger (more careful) observation
leading to faster (more complete) dephasing.

\begin{figure}\label{fig2}
\begin{center}
\includegraphics*[width=8cm,height=10cm,keepaspectratio]{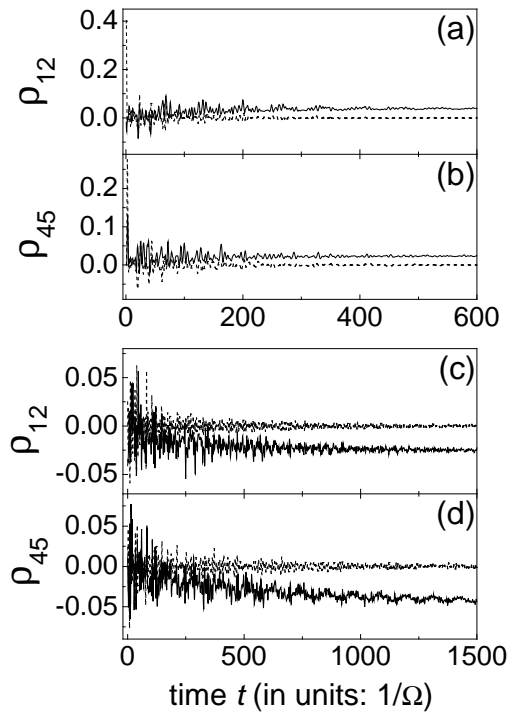}
\caption{Measurement-induced dephasing behaviors in the presence of inelastic tunneling
through the QPC, in low voltage regime ($V=1.5\Omega$). Here, by the solid and dashed
curves, we plot the real and imaginary parts of the off-diagonal matrix elements
$\rho_{ij}$, which are defined in the local well-state basis.
Distinct dephasing rates are obtained from two different models: the faster dephasing
behaviors in (a) and (b) correspond to stronger measurement in attempt to distinguish the
electron's position in each quantum well; the much slower ones in (c) and (d) are
resulted from observation that only determines the electron in the first (nearest)
quantum well.     }
\end{center}
\end{figure}

Note that in the local well-state representation, the real
part of the off-diagonal matrix element does not vanish after
sufficient measurement, e.g., see Fig.\ 2(a).
The underlying reason is that the back-action of the present continuous
weak measurement plays a role of an {\it effective thermal bath},
in particular the measurement voltage is equivalent
to an {\it effective temperature} \cite{Li04}.
As a result, the long-time (sufficient) measurement will make the
measured state be an {\it effective thermal state}, leading to zero
off-diagonal elements of the density matrix between the eigen-energy states.
Thus, in general the density matrix is not diagonal in the alternative
quantum-well-state basis, where the {\it coherence} between the
quantum-well-states originates from the superposition components
(branch waves) of the same {\it eigen-energy state}, but not between them.
In Ref.\ \onlinecite{Gur97} and \onlinecite{Gur00}, the high voltage limit
leads to an equal occupation probability in the effective thermal state,
which makes the density matrix be diagonal in arbitrary state basis.

Another consequence of back-action is the measurement-induced
relaxation, as numerically shown in Fig.\ 3, where the adopted
parameters are the same as above.
Rather than the restriction to the
large voltage limit as in Ref.\ \onlinecite{Gur00}, here we
particularly focus on the low voltage regime, say, $V<\Delta$,
with $\Delta$ the disorder strength.
Initially (at time $t=0$), the electron is assumed in the ground
state, with a distribution probability dominantly localizing in
the eighteenth well, as shown by the solid curves in Fig.\ 3. As a
result of the measurement, the state relaxation gradually takes
place, i.e., {\it delocalization} leads to re-distribution of
electron probability in each well.
Note that our result shown in Fig.\ 3(a) differs considerably from that
in Fig.\ 3(b) based on the Lindblad-type master equation \Eq{LB-ME}.
The latter shows that after sufficient relaxation each well is
occupied with identical probability, which was indeed
proven {\it analytically} in Ref.\ \onlinecite{Gur00}.
However, our treatment leads to un-equal occupation
probabilities in each well.
This discrepancy originates from whether
the inelastic energy exchange between the detector and the
measured system is properly included in the transition rates \cite{Li04,Li05},
which leads to a non-trivial detailed balance condition.
Remarkably, we notice that ignoring this inelastic effect
in the transition rates is equivalent to
assuming the large voltage limit.
This is in particular illustrated by Fig.\ 3(c) in comparison with Fig.\ 3(b).

{\it Output Current}.---
The un-conditional dynamics described by \Eq{ME-3} appropriately
addresses the measurement induced back-action. However,
measurement information cannot be drawn out from it. In this
sense, the conditional version of master equation (\ref{ME-2})
contains rich measurement information. Based on it, one can study
the readout characteristics of the detector, such as the output
noise spectrum and current \cite{Li05}.
Here we exemplify how to calculate the measurement current. From
$\rho^{(n)}(t)$, the distribution function reads
$P(n,t)=\mb{Tr}[\rho^{(n)}(t)]$, which describes the probability
of finding ``$n$" electrons tunnelled through the QPC barrier up
to time ``$t$". As a result, the measurement current is obtained
as \cite{Li05} \bea\label{it} I(t)&\equiv& e \frac{dN(t)}{dt} = e
\sum_n n \mb{Tr}[\dot{\rho}^{(n)}(t)]   \nl
     &=& \frac{e}{2}\mb{Tr}[\bar{Q}\rho Q + \mb{H.c.} ] ,
\eea where $\bar{Q}\equiv \ti{Q}^{(-)}-\ti{Q}^{(+)}$. In
particular, the stationary measurement current can be expressed in
general as
\begin{equation}\label{Iss}
I_{ss}=I^{(0)}+\tilde{I} .
\end{equation}
Here $I^{(0)}=\sum_{j=1}^{N}I_j\rho_{jj}$, with $\rho_{jj}$ the
occupation probability of the $j_{\rm th}$ well, and $I_j$ the
associated readout current. Explicitly, $I_j=Vg_j$, with $g_j$ the
conductance defined by $g_j=e^2\eta(\Omega_0+\chi_j)^2$.
As found in Ref.\ \onlinecite{Li04}, long-time sufficient
measurement only results in complete dephasing between
eigenstates, but not the local quantum-well states, i.e.,
$\rho_{jj'}\neq 0$, which leads to the non-vanishing term
$\tilde{I}$ in \Eq{Iss}.

As an explicit example, consider a two-state system (qubit). Let
us denote the local energy-level offset by
$\epsilon=(\epsilon_2-\epsilon_1)/2$, and $\Delta=
E_2-E_1=2\sqrt{\epsilon^2+\Omega^2}$ for the eigen-energy
difference. For convenience, we also introduce
$\sin\theta=2\Omega/\Delta$.
At zero temperature, for low bias voltage $V<\Delta$, the
stationary current is
$I_{ss}=I_1\rho_{11}+I_2\rho_{22}+\frac{\sin\theta}{2}(\chi_1-\chi_2)^2\eta
V\rm{Re}\rho_{12}$. For voltage $V>\Delta$, the current reads
$I_{ss}=I_1\rho_{11}+I_2\rho_{22}+\frac{\sin\theta}{2}(\chi_1-\chi_2)^2\eta
\Delta\rm{Re}\rho_{12}$.
Here, for qubit measurement, we have analytically shown the
current ``correction" due to the non-vanishing $\rho_{12}$.

\begin{figure}\label{fig3}
\begin{center}
\includegraphics*[width=8cm,height=10cm,keepaspectratio]{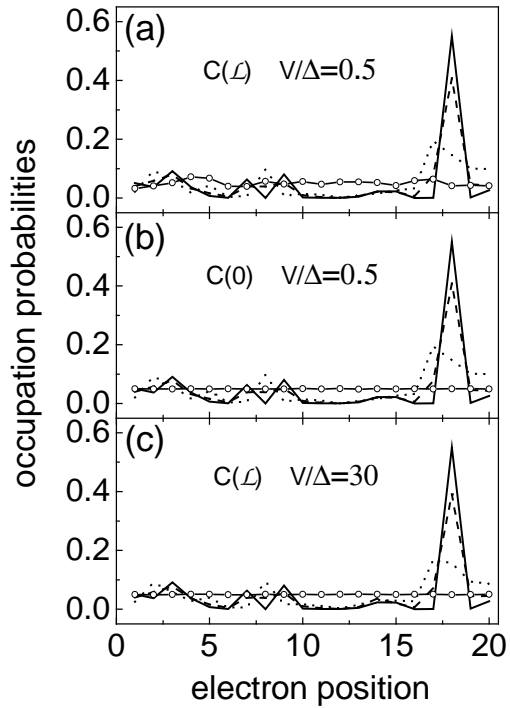}
\caption{ Measurement-induced delocalization of the electron, initially which is
dominantly localized in the eighteenth quantum well (in the ground state), as denoted by
the solid curves.
Shown in the figure by the dashed, dotted, and symbolized curves are the distribution
probabilities in each well, at times $0.4\Omega^{-1}$, $0.8\Omega^{-1}$, and
$500\Omega^{-1}$, respectively.
In low voltage regime, the detailed-balance preserved result in (a) differs considerably
from that in (b) obtained from the Lindblad master equation \Eq{LB-ME}. At high voltage
limit, the result in (c) from \Eq{ME-3} recovers the prediction by \Eq{LB-ME}. }
\end{center}
\end{figure}

In the following, we study numerically the characteristics of
measurement current for more complicated (multi-state) system.
To be definite, consider an array of four coupled quantum wells,
which is parameterized by $\epsilon_j/\Omega=2,4,1,0$ (j=1,2,3,4).
The corresponding eigen-energies read $E_1=-0.69\Omega$, $E_2=-1.2\Omega$,
$E_3=1.83\Omega$, and $E_4=4.67\Omega$.
Also, let us denote the eigen-energy differences by $\Delta_{ij}=E_i-E_j$.
In Fig. 4(a) we found three transition voltages,
$V_1=1.9\Omega$, $V_2=2.55\Omega$, and
$V_3=3.5\Omega$, which correspond to, respectively,
the transition energies $\Delta_{21}$, $\Delta_{31}$, and $\Delta_{42}$.
The conductance plateaus in Fig. 4(a) can be accordingly
understood as follows.
In the low voltage regime, $V<V_1$, the electron relaxed onto the ground
state that contains the dominant components of local well states
$|4\ra$ and $|3\ra$. The differential conductance is thus approximated by
$g_3\rho_{33}+g_4\rho_{44}$.
In the regime $V_1<V<V_2$, the detector back-action causes
re-excitation of electron from the
ground state $|E_1\ra$ to $|E_2\ra$, and from $|E_2\ra$ to $|E_3\ra$.
These processes lead to the obvious changes of occupation
probabilities on local states $|4\ra$ and $|1\ra$ [Fig. 3(c)],
which result in the first conductance plateau in Fig. 3(a).
If $V>V_2$, direct excitation from $|E_1\ra$ to $|E_3\ra$ takes place,
leading to jumps of the local state probabilities $\rho_{11}$ and
$\rho_{44}$ as shown in Fig. 3(c), and the second conductance
plateau in Fig. 3(a).
Finally, as $V>V_3$, excitation from $|E_2\ra$ to $|E_4\ra$ is induced
by the back-action. As a consequence, the occupation probability
of the local state $|2\ra$ has a jump, since $|2\ra$ is
the dominant component of the eigenstate $|E_4\ra$. This leads to
the last observable conductance plateau in Fig. 3(a).
\begin{figure}\label{fig4}
\begin{center}
\includegraphics*[width=9cm,height=10cm,keepaspectratio]{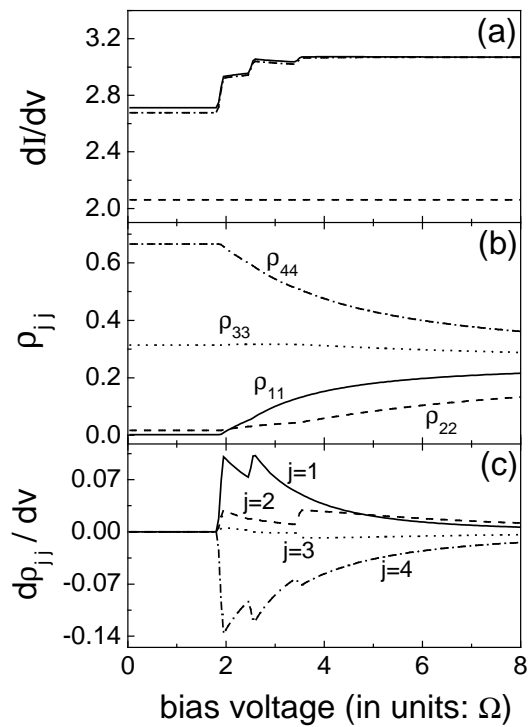}
\caption{ (a) Measurement current (plotted in terms of the non-linear differential
conductance $dI/dV$) versus the bias voltage across the QPC detector. Here the solid,
dot-dashed and dashed curves are, respectively, resulted from $``I^{(0)}+\tilde{I}"$,
$``I^{(0)}"$ and Gurvitz's Bloch equation approach \cite{Gur97,Gur00}. (b) The associated
occupation probabilities in the individual local wells, and (c) their differential
altering rates with the voltages. Based on (b) and (c) together with the current formula
\Eq{Iss} the conductance plateaus in (a) can be readily understood (see the main text for
details). }
\end{center}
\end{figure}

{\it Concluding Remarks}.--- Based on the detailed-balance
preserved \Eqs{ME-2} and (\ref{ME-3}), we have studied the quantum
measurement problem of a multi-state system measured by the
mesoscopic QPC detector.
The implication of present study is twofold. On one hand, via this
multi-state model system, we illustrated that in general the QPC
measurement setup cannot be described by the Lindblad-type
master equation [i.e. \Eq{LB-ME}], which has been commonly assumed
(either explicitly or implicitly)
in recent literatures \cite{Gur97,Gur00,Goa01,Kor99}.
This conclusion may
have impact on the future study of solid-state quantum measurement
and quantum feedback control.
On the other hand, in comparison with previous studies on the
measurement-induced delocalization, particularly Ref.\
\onlinecite{Gur00}, we shed new light on the underlying physics.
Namely, rather than owing to the mere loss of quantum coherence
\cite{Gur00}, the delocalization stems largely from the energy
exchange of the measured electron with the measurement device QPC,
thus depends on the measurement voltages and leads to a number of
distinct delocalization features.

\vspace{2ex} {\it Acknowledgments.} Support from the National
Natural Science Foundation of China, and the Major State Basic
Research Project No.\ G001CB3095 of China are gratefully
acknowledged.


\clearpage

\end{document}